\documentclass[11pt]{article}
\usepackage{authblk}
\usepackage{hyperref}
\usepackage{natbib}
\usepackage{graphicx}
\usepackage{verbatim}
\usepackage{placeins}

\usepackage[vscale=0.88,hscale=0.8,includehead,vmarginratio=1:9]{geometry}
\usepackage{times}

\begin{document}

	\title{Calibrating GONG Magnetograms with End-to-end Instrument Simulation I: Background, the GONG Instrument, and End-to-end Simulation}
	\author[1]{Joseph Plowman}
	\author[2]{Thomas Berger}
	\affil[1]{National Solar Observatory, \url{jplowman@nso.edu}} 
	\affil[2]{University of Colorado at Boulder}
	\date{}
	\maketitle
	\begin{abstract}
		This is the first of three papers describing an `absolute' calibration of the GONG magnetograph using an end-to-end simulation of its measurement process. The input to this simulation is a MURaM 3D MHD photospheric simulation and the output is the corresponding set of simulated data numbers which would be recorded by the GONG detectors. These simulated data numbers are then used to produce `synthetic magnetograms' which can be compared with the simulation inputs. This paper describes the GONG instrument, the MURaM datacube, our instrument simulator, and calculation of synthetic magnetograms, setting the stage for the subsequent two papers. These will first lay groundwork for calibration (and magnetogram comparison in general), then apply them to calibration of GONG using the simulation results. 
	\end{abstract}

	\section{Introduction}\label{sec:intro}

	Measurement of the solar photospheric magnetic field is well-established and long-standing: routine observations have been now been made for over 75 years \citep[see][]{Babcock53, HowardBabcock60, HowardEtal83}. A variety of instruments now regularly measure the photospheric magnetic field: some have high resolution and narrow field of view designed for close-up observation of varying solar features of interest (e.g., the space-based {\em Hinode}/Solar Optical Telescope, SOT, or the ground-based Swedish Solar Telescope, SST). Others are {\em synoptic} instruments, designed to monitor the entire solar hemisphere visible from Earth. Ground-based examples include the National Solar Observatory (NSO) Global Oscillations Network Group \citep[GONG;][]{HarveyEtal_GONG1996} and Synoptic Optical Long-term Investigations of the Sun/Vector Spectromagnetograph \citep[SOLIS/VSM;][]{KellerEtal_SOLIS2003}. Space-based examples of synoptic magnetographs include the Solar Dynamics Observatory/Helioseismic and Magnetic Imager \citep[SDO/HMI;][]{ScherrerEtal_HMI2012} and Solar and Heliospheric Observatory/Michelson Doppler Imager \citep[SoHO/MDI, which suspended operations in 2012;][]{ScherrerEtal_MDI1995}.

\begin{figure}
		\begin{center}\includegraphics[width=\textwidth]{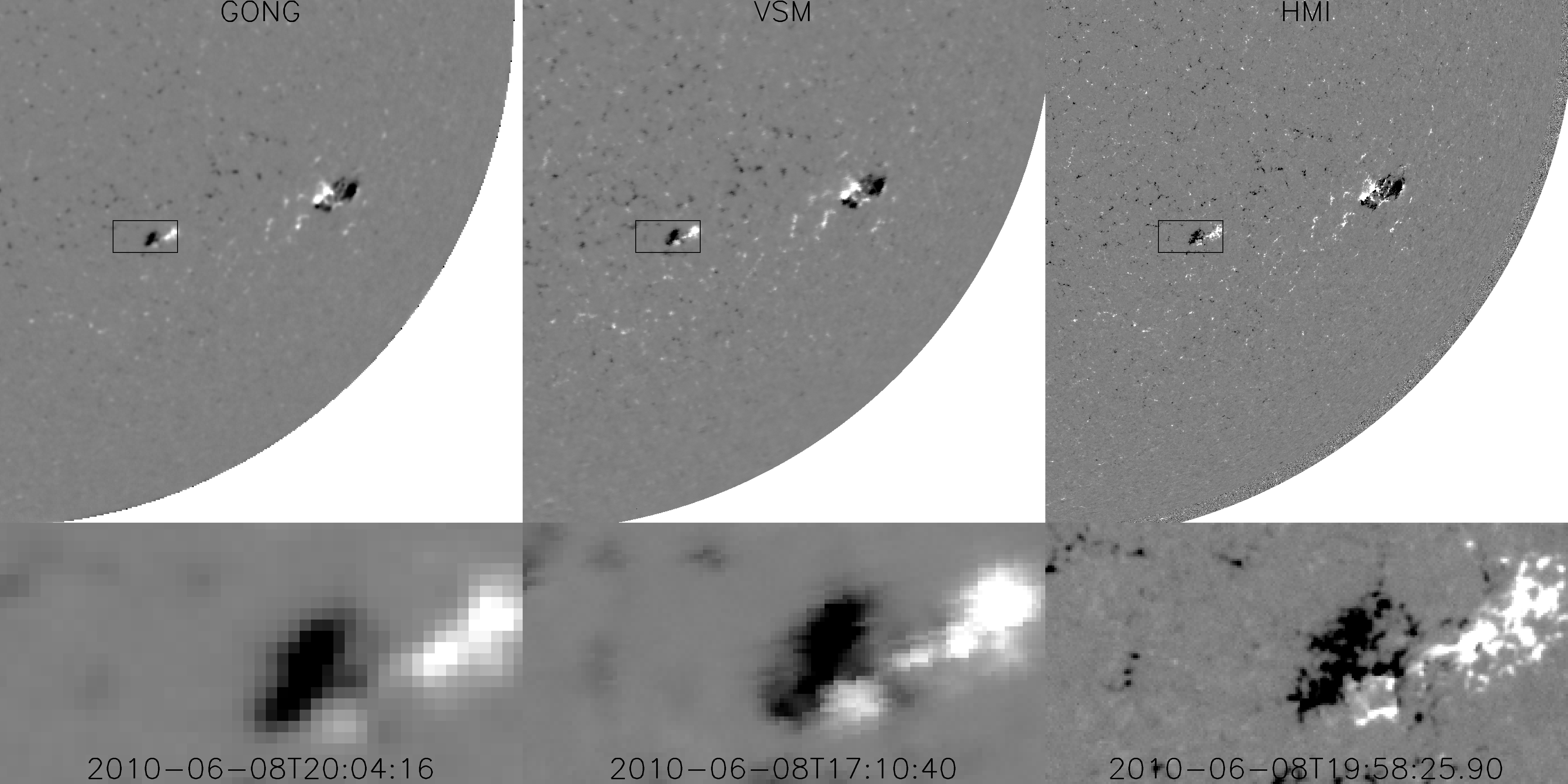}\end{center}
		\caption{Examples of magnetic field measurements from GONG (left), VSM (center), and HMI (right). Intensity represents line-of-sight magnetic field strength ranging from -150 Gauss (black) to 150 Gauss (white).}\label{fig:examplemagnetograms}
	\end{figure}

	The subject of the work reported in this paper is a new `absolute' calibration of the GONG synoptic magnetograph. GONG, along with SOLIS/VSM, SDO/HMI, and (where long-duration observations are needed) the low-resolution Wilcox Solar Observatory \citep[see][for example]{HoeksemaScherrer1986} are the most commonly used syoptic magnetographs (See Figure \ref{fig:examplemagnetograms} for example magnetograms from GONG, VSM, and HMI). This work is motivated by the importance of GONG measurements for space-weather, a variety of apparent differences between the measurements and {\em in situ} field measurements (and between the measurements themselves), and a lack of absolute calibration of the measurements. We begin by reviewing these issues.

	A primary use of synoptic magnetograms is as the `boundary condition' for models of the coronal magnetic field. These, in turn, are the primary inputs to space weather forecasts of the solar wind at Earth. Solar wind forecast models such as WSA/Enlil \citep[][used in operations by NOAA, the US Air Force (USAF), and the UK MetOffice]{ArgePizzo_JGR2000}, provide current predictions of geomagnetic storms due to the solar wind and Coronal Mass Ejections (CMEs). Space weather models almost all use these photospheric field measurements in this way, making them the foundational inputs to their predictions: all of the other modeling steps depend on them, so any issues with the measurements will propagate throughout the models and resulting predictions. One such issue is that the measurements have not been calibrated in any absolute sense, owing to the lack of `ground truth' measurements to complement them -- e.g., {\em in situ} magnetometer measurements {\em in} the photosphere; these are not physically possible for obvious reasons.

	This lack of a magnetogram calibration baseline is well illustrated by the wide divergence of results found in studies comparing full-disk magnetograms \citep[e.g.,][]{VirtanenMursula2017, 2013PietarilaEtal_SoPh282_91,2012LiuEtal_SoPh279_295L}, synoptic maps constructed from various instruments \citep[e.g.,][]{Riley_comparison2014}, and higher-resolution, limited field-of-view, magnetograms \citep{2002BergerLitesI_SoPh213_213B, 2003BergerLitesII_SoPh213_213B}. Many of these studies find significant apparent differences in magnetic flux values measured with different instruments. \cite{2012LiuEtal_SoPh279_295L} compare SDO/AIA with SoHO/MDI and find that the lower-resolution MDI measurements are {\em higher} than the HMI's by a factor of $1.4$. \cite{Riley_comparison2014} compare seven synoptic maps created from different input magnetograms and find that their magnetic flux levels can differ by an order of magnitude. \cite{VirtanenMursula2017} compare measurements from six full-disk magnetograms, analyzing their data in terms of a harmonic expansion, and find considerably better agreement than some other synoptic map comparisons, with levels between SDO/HMI and SOLIS/VSM reasonably approximated by a factor of 0.8. \cite{2002BergerLitesI_SoPh213_213B} find a factor of 0.6 for a single sunspot active region using the highest spatial resolution magnetograms obtained at that time from the SST and the SOHO/MDI ``high-res'' measurements. 

	Further reinforcing the need for absolute magnetogram calibrations, the `open' flux (i.e., the amount of flux that reaches from the photosphere, through the corona, and into interplanetary space to be measured at 1 AU) estimated from {\em in situ} measurements \citep[e.g.,][]{1998StoneEtal_ACE_SSRv86_1S, 1995LeppingEtal_WindMFI_SSRv71_207L}) are at least two times higher than model predictions extrapolated from the photospheric magnetic field measurements \citep{LinkerOpenFlux2017}.
	Without a reliable baseline magnetogram calibration to use, it has become common practice among solar wind modelers to simply multiply any given synoptic magnetic field map by a factor of two or even four \citep[MacNeice, private communication; see][who provide a number of references documenting this practice]{LinkerOpenFlux2017, Riley_comparison2014} as a way to force the extrapolations to match the {\em in situ} field measurements at 1 AU.
	
	One hypothesis is that these differences are due purely to the resolution differences between the instruments. \cite{2002BergerLitesI_SoPh213_213B}, for example, point out that spatial resolution differences between instruments can be a major source of erroneous inter-calibration, because the smallest-scale bipolar magnetic structures are ``cancelled out'' in the lower resolution instrument measurements in a way that cannot be inverted for comparison to higher spatial resolution instruments. In other words, higher spatial resolution instruments would be expected to measure higher {\em  minimum and maximum} values of the  magnetic {\em field} in small-scale structures outside of sunspots, since they are able to spatially distinguish the ``salt and pepper'' positive and negative polarity mixtures found in these regions. 
	
	This alone would {\em not} explain why the instruments themselves would measure different {\em fluxes}, however: Like the fluxes, the measurements are area-integrated quantities. Therefore, in order to compare {\em fluxes} from two different instruments one must ensure they have been integrated over the same area (otherwise they aren't commensurate; they don't have the same units). These `salt \& pepper' differences will cancel out in the area integration, assuming both instruments' field measurements are otherwise identical: aye, there's the rub!

	Lack of absolute magnetograph calibration almost certainly plays a role in this open flux problem, but it is not alone. Other issues are determining the radial field from line-of-sight observations, the fact that the real coronal field is not potential, and the limitations of the source surface assumption. Perhaps the most significant is the limitation of current magnetograms to Earth's perspective versus the global nature of the coronal magnetic field: Flux concentrations (e.g., Sunspot active regions) on one hemisphere of the Sun can have a major influence on coronal magnetic topology in the other hemisphere, particularly in the global heliospheric current sheet. Therefore, accurate modeling of the coronal magnetic field, and hence the solar wind at any position in the heliosphere, requires knowledge of the magnetic field {\em over the entire solar photosphere}; the sun's poles are especially important.
	
\begin{figure}
		\begin{center}\includegraphics[width=0.6\textwidth]{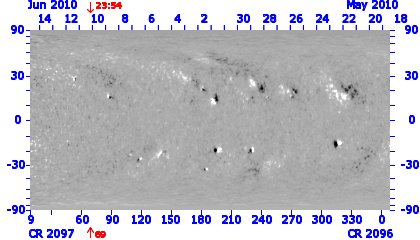}\end{center}
		\caption{Example GONG synoptic map for Carrington rotation 2097, which includes the time of the example magnetogram snapshot shown in Figure \ref{fig:examplemagnetograms}.}\label{fig:example_synopticmap}
	\end{figure}

	The unfortunate reality of our current Earth-bound perspective forces coronal magnetic field and solar wind modelers to make more or less erroneous assumptions to construct so-called `synoptic maps' of the global photospheric magnetic field for their boundary conditions. In addition to the lack of far-side observations, the fact that the Earth is in the ecliptic means our view of the poles, even when they are visible, are seen at a very high inclination, which makes accurate measurements difficult or impossible. As a result, synoptic maps must fill in the highest latitudes using some sort of extrapolation. Figure \ref{fig:example_synopticmap} shows a GONG synoptic map of the entire solar sphere for Carrington Rotation 2097, created by assuming that the magnetic structures measured on the Earth-facing hemisphere simply rotate around the far side of the sphere.
	
	More sophisticated synoptic maps \citep[e.g., ADAPT:][]{2015HickmannEtal_SoPh290_1105H} may attempt to correct for many of these effects: solar differential latitudinal rotation rates, the latitudinal shear of active regions as they rotate around the Sun, extrapolation of the polar fields based on the known pole-ward migration of surface flux over time, and so on. However, they still cannot account for any active regions that might emerge on the far side of the Sun. Nevertheless, these GONG synoptic maps are the primary boundary condition inputs to most current solar wind forecasting models, including WSA/Enlil. 
	
	This is the first of three papers in which we tackle the absolute calibration of the GONG full-disk magnetograms. Our objective is provide a calibrated  magnetic flux measurement that is suitable for solar wind prediction models. There is no reason to expect {\em a priori} that {\em all} of the discrepancy between solar wind models and the {\em in situ} measured magnetic flux at 1 AU is due to calibration errors in photospheric magnetograms, and {\em none} come from the other issues. However, the photospheric field measurements are the foundational inputs to these predictions: all of the other steps depend on them, so it is certain that undertaking the magnetogram calibration effort described here will advance the fields of solar magnetic field measurement and space weather forecasting. 

	We solve the problem with an `end-to-end' simulation of the GONG measurement process: A MURaM 3D MHD simulation \citep[MURaM;][]{Rempel2015} provides a numerical `ground truth' (magnetic field, temperature, density, and velocity). From the MURaM simulation, we produce spectral cubes using a Rybicki and Hummer (RH) radiative transfer model \citep{Uitenbroek_ApJ2001}. We then simulate all of the major physical and numerical processes that comprise a full-disk GONG magnetogram observation. By feeding this instrument simulation with the RH spectral cubes, we can determine the input (`ground truth') magnetic flux level (from the MURaM simulation) for every output (`measured') magnetic flux (the output of the GONG simulation). Comparison between the two produces a set of calibration curves for GONG magnetograms that can be used to correct both full-disk magnetograms and the synoptic maps that are created from them. 
	
\begin{figure}
		\begin{center}\includegraphics[width=\textwidth]{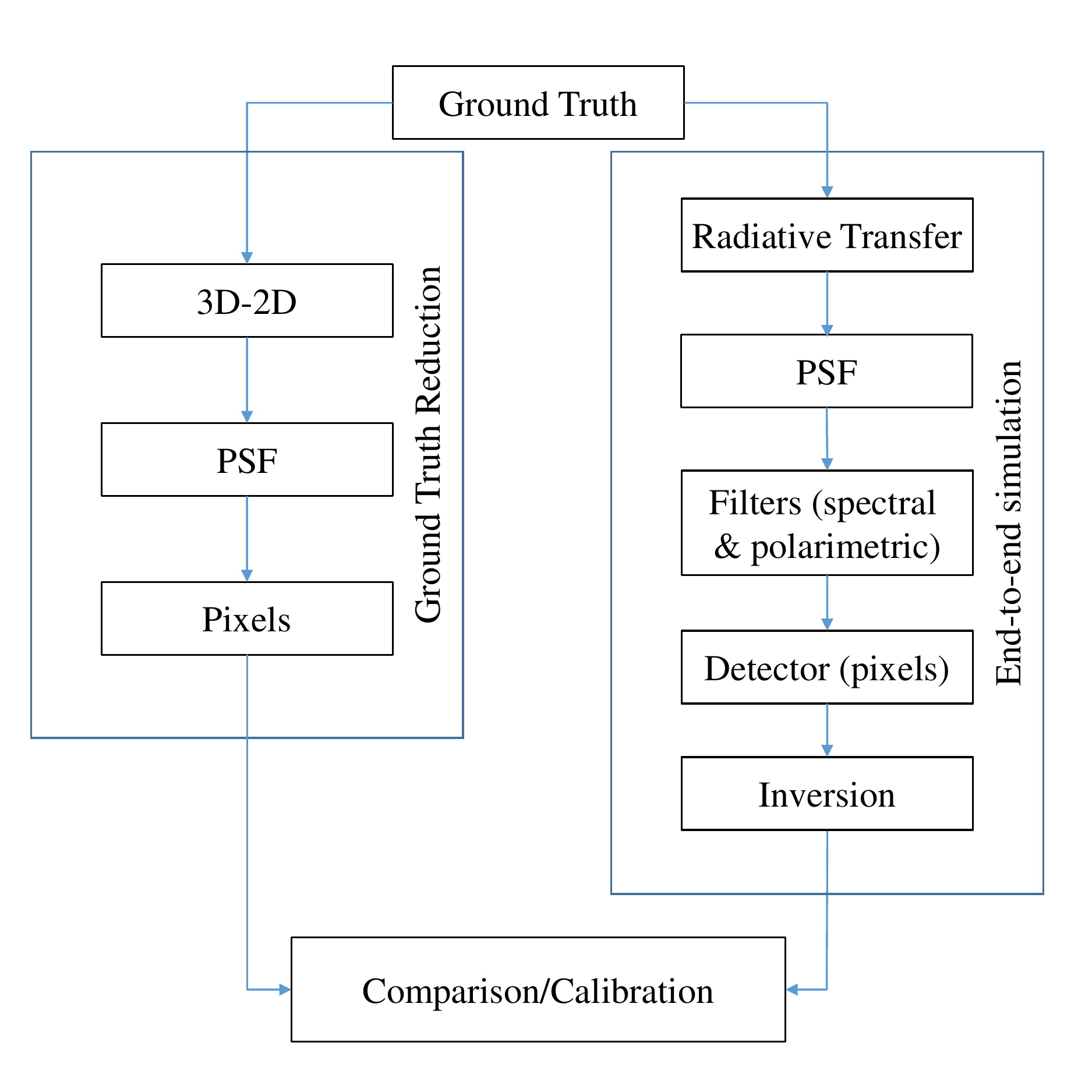}\end{center}
		\caption{Flow chart of the calibration process. On the right is the end-to-end simulation of the measurement process, which we describe in this paper, while on the left is the `ground truth reduction' which places the ground truth in a form that can be directly compared with the measurements \citep[described in the next paper of the series][]{PlowmanEtal_2019II}.}\label{fig:calibration_boxdiagram}
	\end{figure}

	Our approach is illustrated in Figure \ref{fig:calibration_boxdiagram}. On the right hand side of the figure is the radiative transfer and instrument simulation, while on the left is the process by which the ground truth is reduced to make a `perfectly' calibrated version of the measurements, suitable for use as a calibration. This first paper \citep{PlowmanEtal_2019I} focuses on describing the GONG instrument and our simulation of it (the right-hand side of Figure \ref{fig:calibration_boxdiagram}). In Sec.~\ref{sec:GONGprinciple} we describe the GONG instrument and magnetic field measurement process in detail. Section~\ref{sec:endtoend} describes our `end-to-end' simulation of the creation of simulated magnetograms using the MURaM simulation and RH radiative transfer code, while Section \ref{sec:summary} shows examples of the simulation results and provides some closing thoughts.

	In the subsequent papers \citep{PlowmanEtal_2019II, PlowmanEtal_2019III}, we will explain the left-hand side of Figure \ref{fig:calibration_boxdiagram} and the comparison/calibration results. In the process, we will investigate some major issues present in all calibrations or inter-calibrations of solar magnetograms and apply the lessons learned to the specific case of calibrating GONG magnetograms against a numerical simulation ground truth. We will also discuss some of the general findings that can be applied to other magnetogram calibration efforts. These papers are the first phase of a research project funded under a NASA space weather `operations to research' (O2R) grant whose specific goal is to improve the accuracy of solar wind prediction models.


	\section{GONG Instrument Description and Measurement Algorithms}\label{sec:GONGprinciple}


	The papers describing the GONG instrument are sporadic and now rather dated \citep[e.g.,][]{HarveyEtal_AdSpR1988, HarveyEtal_GONG1996}, so this section gives an overview of its basic longitudinal magnetograph and dopplergraph operating modes. This understanding is based on a review of NSO internal documentation and discussion with NSO staff involved in its construction and operation, and forms if the basis of our GONG instrument simulation.

	GONG consists of 6 identical installations at sites around in the world: Big Bear, Mauna Loa, Learmonth, Udaipur, Teide, and Cerro Tololo. The purpose of this distribution is to allow continuous solar observation from the ground despite nightfall and cloud cover, and the original design focus was Doppler imaging for helioseismology. Magnetic field measurement was only a minor design driver, but that capability has become an increasingly important usage of GONG, especially for space weather prediction: of the papers on NASA ADS citing the primary GONG instrument paper \citep[GONG;][]{HarveyEtal_GONG1996} which use the GONG magnetograms, over 75\% use them for space weather global field extrapolations. 

	Each of the 6 GONG instruments are housed in a cargo container with dimensions $2.4\times 2.6\times 6.1$ meters, which serve as both shipping containers and on-site instrument shelters. Light is fed into the instrument by a movable sealed turret assembly, consisting of a (slightly wedged and tilted) entrance window with a heat-reflecting dielectric coating on the inside surface followed by two low-polarizing dielectric mirrors at 45 degree incidence angle. This system lets through light in a $\sim 50$ nm bandpass centered on 677 nm, which is then focused by an achromatic lens with a 1 meter focal length to a $\sim ~10$ mm solar image.

	Close to the focal plane is a liquid crystal variable retarder (LCVR) that is set to either 1/4 or 3/4 wave retardation. It is followed by a polarizing beam splitter cube that transmits light at 677 nm (both $p$ and $s$) straight through, but reflects $s$ - light ($s$ refers to light polarized perpendicular to plane of incidence, while $p$ - light is polarized parallel to the plane) at 656 nm to the auxiliary H-alpha system. Next in line is a relay lens and an ion-assisted deposition interference filter in a temperature controlled oven that transmits a $\sim 0.5$ nm band pass centered on 676.78 nm. The form of this spectral response function can vary from filter to filter; an `ideal' form can be written 
	\begin{equation}
		R_p = \big[1+(\lambda-\lambda_0)^4/(2w_p)^4\big]^{-1},
	\end{equation}
	where $w_p=0.5$ nm is the filter FWHM and $\lambda_0=676.78$ nm is the central wavelength. A closer to real-world version, found in GONG instrument memos, is
	\begin{equation}\label{eq:GONG_prefilter_transmission}
		R_p = \Bigg[1+1.6569\Big(\frac{\lambda-\lambda_0}{w_p}\Big)^2\Bigg]^{-2}.
	\end{equation}
	We checked for differences between these functions and found they have negligible effect on the magnetograms ($\sim 0.15\%$). In our simulations, we have used the second of these expressions (Equation \ref{eq:GONG_prefilter_transmission}).

	Next comes another oven containing a 3-element Lyot filter. In front of the Lyot, the oven also contains a polarizing beam splitter cube which sends $s$ - light to a guider and transmits $p$ - light into calcite crystals. At this point the solar polarization analysis of the instrument is finished since only originally left or right circularly polarized light (depending on LCVR retardation) passes further into the filter. The Lyot filter has a free spectral range of about 0.8 nm and a passband of about 0.1 nm centered on the 676.8 Ni I line -- its spectral response can be written

	\begin{equation}
		R_L = \cos^2{\Big(\frac{\pi \Delta h}{\lambda}-\theta_1\Big)}\cos^2{\Big(\frac{2\pi \Delta h}{\lambda}-\theta_2\Big)}\cos^2{\Big(\frac{4\pi \Delta h}{\lambda}-\theta_3\Big)},
	\end{equation}
	where $\Delta h=608$ microns, $\theta_1 = 67.7$ degrees, $\theta_2 = 45.3$ degrees, and $\theta_3 = 93$ degrees.

	After the Lyot oven is a solid polarizing Michelson interferometer with a path difference between the arms of about 25000 waves. Finally there is a half wave plate, rotating at 5 Hz, synchronized to the camera, which collects frames at 60 FPS. The plate scale of the system is 2.5 arc-seconds per pixel. The effective aperture is 2.8 cm, for which the Rayleigh criterion gives a diffraction-limited spatial resolution of 6.1 arc-seconds (FWHM).

\begin{figure}
		\begin{center}\includegraphics[width=0.5\textwidth]{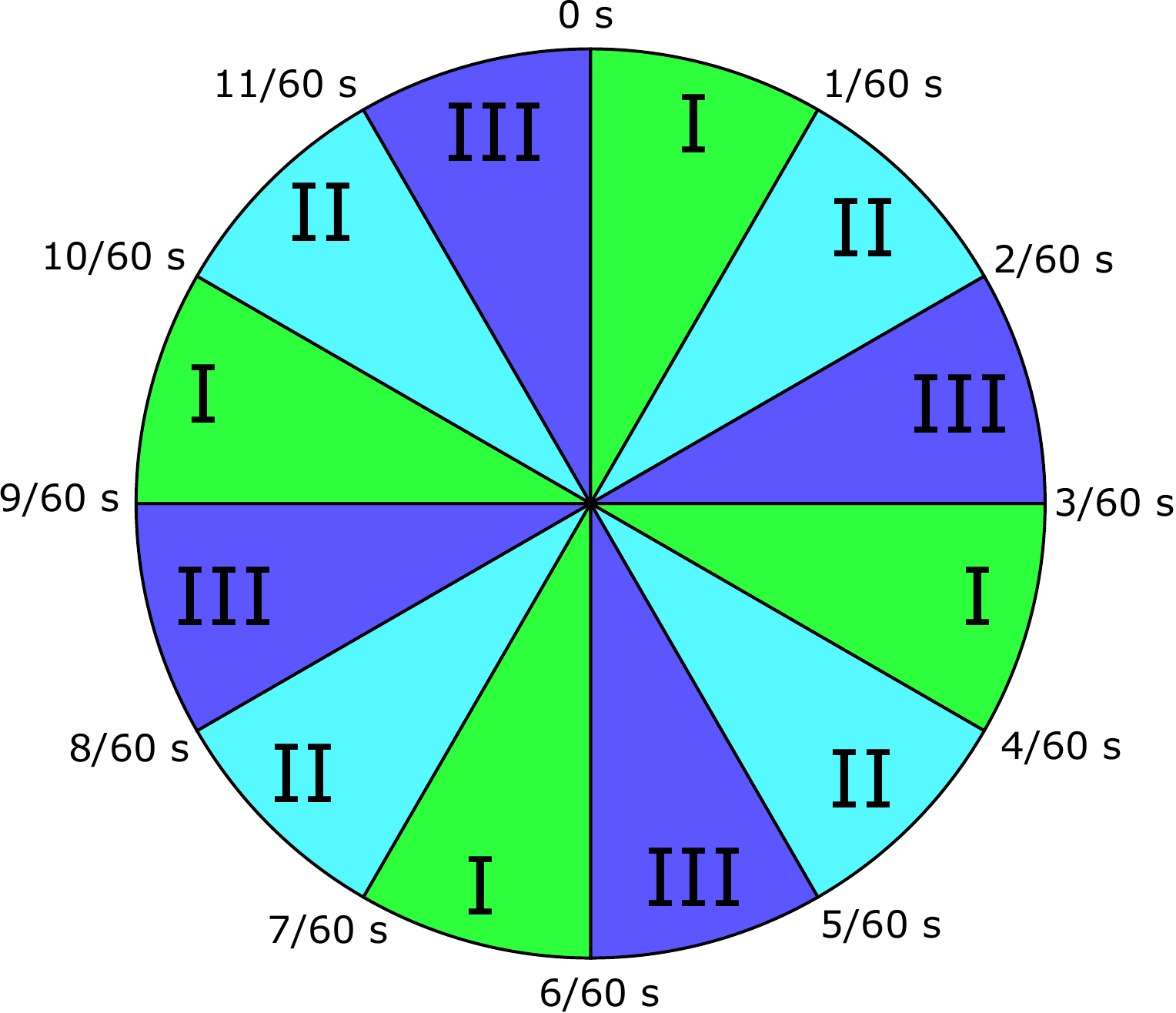}\includegraphics[width=0.5\textwidth]{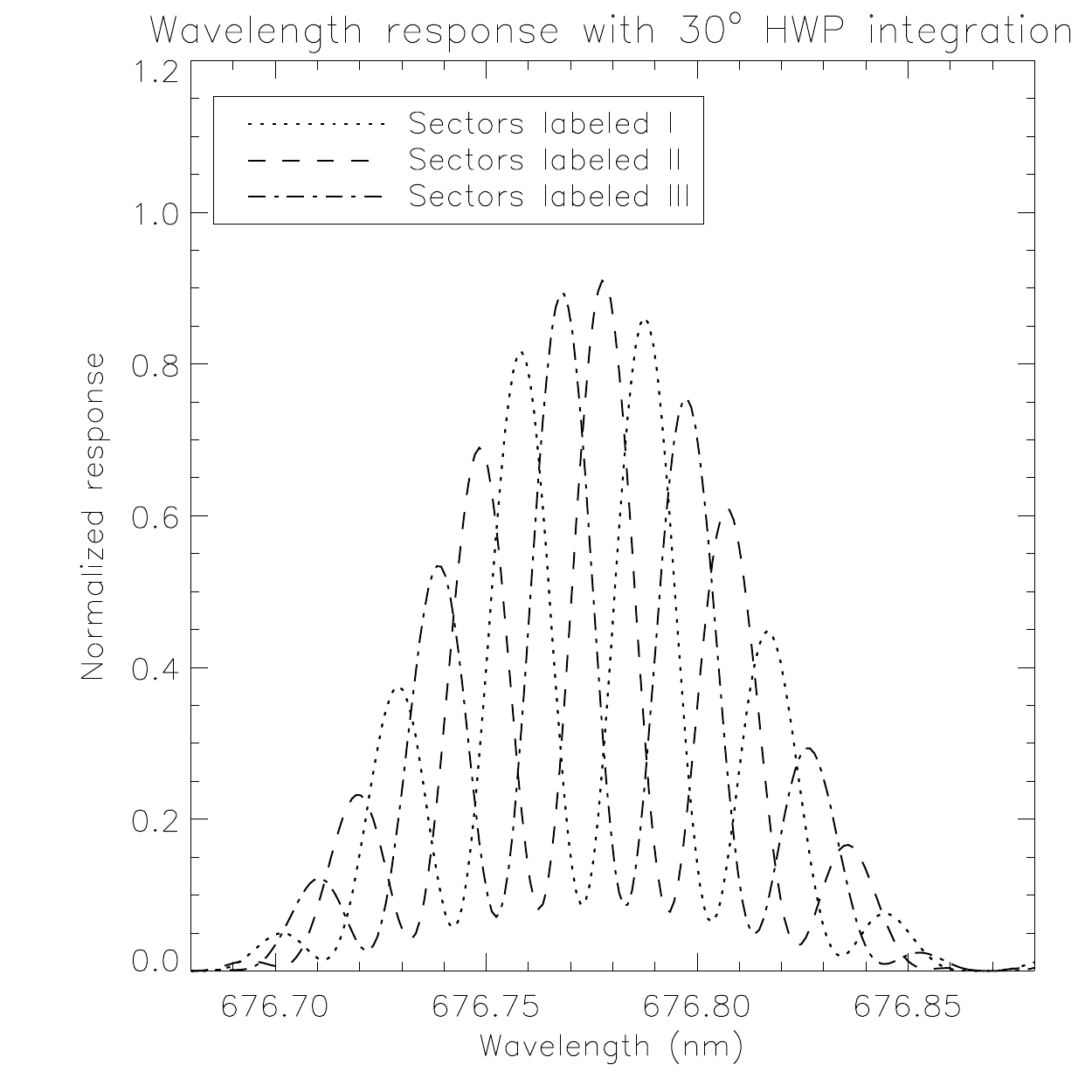}\end{center}
		\caption{Left: Diagram showing how one rotation of the wave plate corresponds to the three unique wavelength response functions in the GONG data. Clock angles in this diagram correspond to the orientation of the waveplate. Right: the three unique GONG wavelength response functions for the 6768 \AA\ line.}\label{fig:GONGclockdiagram}
	\end{figure}

	The channel spectrum of this system, for a given angle of the half-wave plate, can be written

	\begin{equation}
		R_\mathrm{GONG}(\lambda,\theta_w) = R_p R_L \cos^2{\Big(\frac{\pi\lambda_0^2}{\Delta\lambda_m\lambda}+\theta_m-\theta_w\Big)},
	\end{equation}
	where $\Delta \lambda_m=0.305$ \AA\ is the free spectral range (FSR) of the Michelson interferometer, $\theta_m=8.2$ degrees, and $\theta_w$ is the angle of the wave plate with respect to the Michelson reference angle. This angle sweeps the response of the system in wavelength so that, as the waveplate turns, the channel response varies continually across the fixed passband of the Lyot + interference prefilter and the Doppler shifted solar spectral line. Thus, each 1/60 second integration of the GONG cameras is an integration of the channel response function over a 30 degree rotation of the waveplate. The overall response function repeats after a 90 degree rotation of the waveplate, so the camera records 3 distinct channel response images every 20th of a second. Figure \ref{fig:GONGclockdiagram} illustrates this process. Like frames in this sequence are then binned down to produce one set of 3 images for each second of observation. The response of each is then
	\begin{equation}\label{eq:GONGwaveresp}
		R_i(\lambda) = \int_{\frac{i\pi}{6}}^{\frac{(i+1)\pi}{6}} R_\mathrm{GONG}(\lambda,\theta_w)d\theta_w
	\end{equation}
	where $i = 0,1$, or 2: $i=0$ corresponds to sector I in Figure \ref{fig:GONGclockdiagram}, while $i=1$ corresponds to sector II and $i=2$ corresponds to sector III.

	When these response functions are integrated against a spectrum $I(\lambda)$ which has a narrow feature in the middle of the bandpass -- such as the Ni I 676.8 nm line -- the Doppler shift, $\Delta v$, of the line (WRT a reference wavelength for the bandpass) can be inverted as follows:

	\begin{equation}\label{eq:phase_dopplershift}
	\Delta v = v_0\Delta \phi =  v_0\tan^{-1}\Bigg[\sqrt{3}\frac{I_1-I_2}{I_1+I_2-2I_0}\Bigg],
	\end{equation}
	where

	\begin{equation}
		I_i = \int R_i(\lambda) I(\lambda)d\lambda,
	\end{equation}
	and the proportionality constant, $v_0$, between the GONG phase shift and the Doppler shift is $v_0\approx 2151.86$ $\mathrm{m}/\mathrm{s}/\mathrm{rad}$. Note that when these integrals are done numerically, the wavelength range must extend past the range of the prefilter. Otherwise the edges of the wavelength range will affect the inversion.

	Polarization is selected by the LCVR, which changes state every second, so that one set of 3 images (for one second of observation) observes a spectrum for each pixel of one circular polarization state (e.g., Stokes I+V), followed by a second set which observes the complementary state (e.g., Stokes I-V). Thus, a full set of GONG images for the longitudinal magnetograph+dopplergraph mode is acquired every 2 seconds -- three for I+V and three for I-V. We can then write the six GONG observables (one set of six for each pixel) as 

	\begin{equation}\label{eq:GONGintensities}
		I^\mathrm{I\pm V}_i = \int R_i(\lambda)[I_I(\lambda)\pm I_V(\lambda)]d\lambda,
	\end{equation}

\begin{figure}
		\begin{center}\includegraphics[width=\textwidth]{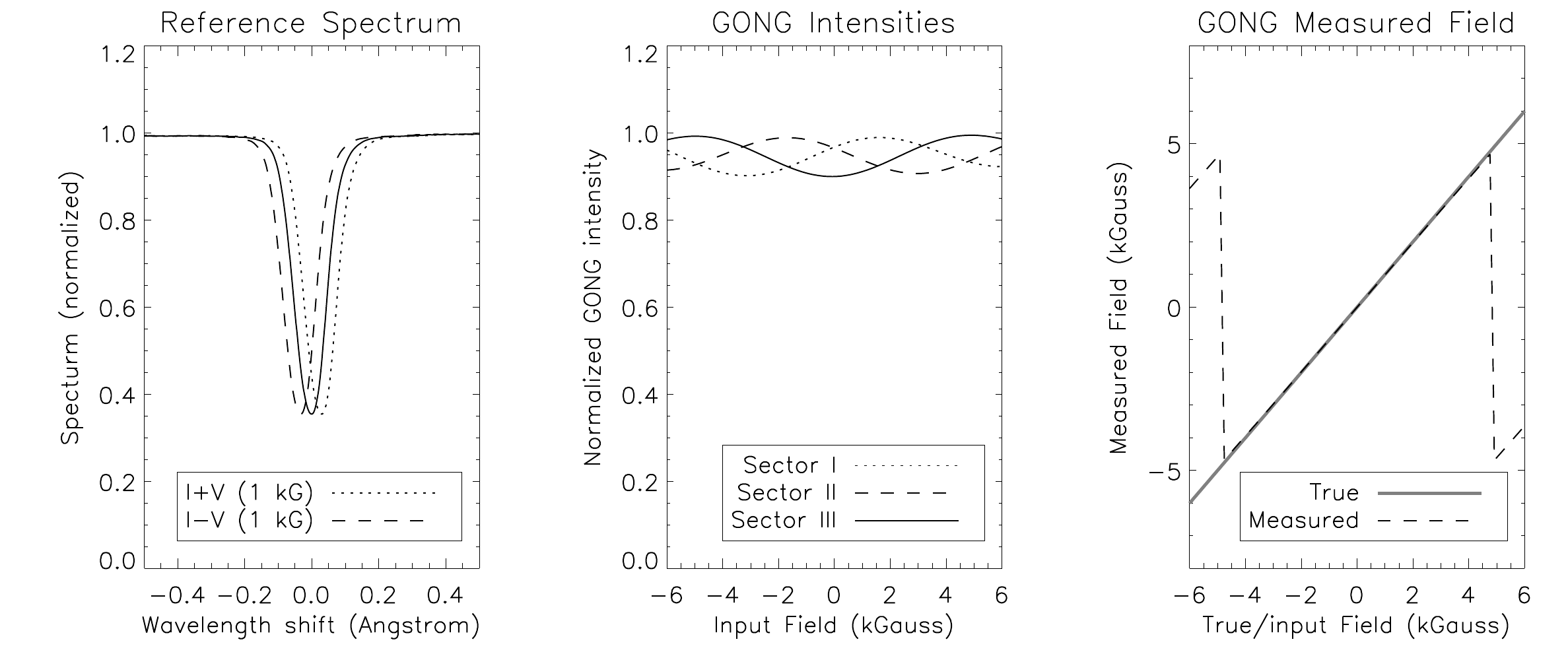}\end{center}
		\caption{Illustration of the GONG magnetic field measurement process. GONG observes solar spectra containing the Ni I 6768 \AA\ line; the presence of a magnetic field causes a wavelength shift in Stokes I+V and I-V (left). The three GONG channel measurements of I+V and I-V vary depending on the wavelength shift, according to Equation \ref{eq:GONGintensities} (center). The measurements are inverted according to Equation \ref{eq:BLOS}, obtaining a measured field strength (right). The linearity of this relationship demonstrates that GONG is not subject to `classical' magnetograph saturation.}\label{fig:GONGfield_measurement}
	\end{figure}

	Where $I_I(\lambda)$ is the spectrum for Stokes I and $I_V(\lambda)$ is the spectrum for Stokes V. Since, to lowest order, the Stokes V profile is proportional to the derivative of the Stokes I profile and Stokes V is small compared to Stokes I, the effect of adding or subtracting Stokes V from Stokes I is to shift the line center to lower or higher wavelength. The magnitude of the field is directly proportional to this shift, and we can write \citep[e.g.,][]{Uitenbroek_ApJ2003}

	\begin{equation}\label{eq:BLOS}
		B_\mathrm{LOS} = \frac{\Delta v^{I+V}-\Delta v^{I-V}}{2}\frac{4\pi m_e}{q_e g_L \lambda_0},
	\end{equation}
	where $m_e$ is the electron mass, $q_e$ is the electron charge, $\lambda_0$ is the rest wavelength of the line, and $g_L$ is its Land\'{e} $g$ factor. For the transition giving rise to the Ni I 676.8 nm line, $g_L$ is nominally 1.5, although a more accurate estimate may be 1.426 \citep{Bruls_AnA1993}. We use $g_L=1.5$ both in the radiative transfer and in the inversion, so that our simulations are self-consistent. In that case, the overall conversion factor is 0.7038 Gauss per m/s of velocity shift (0.03205 \AA\ wavelength shift per kiloGauss).
	This field measurement process is illustrated in Figure \ref{fig:GONGfield_measurement}, which also demonstrates that GONG is not subject to `classical' magnetograph saturation. A very close examination of the curve finds a $\sim 0.5\%$ deviation from linearity over the $\pm 4$ kiloGauss usable range.

	The Doppler shift of the spectral line (apart from the Zeeman effect) is just
	\begin{equation}\label{eq:deltav}
		\Delta v_\mathrm{cen} = \frac{\Delta v^\mathrm{I+V}+\Delta v^\mathrm{I-V}}{2}
	\end{equation}
	and the pseudo-continuum intensity is 
	\begin{equation}\label{eq:pseudocontinuum}
		I = \sum_i I^\mathrm{I+V}_i + I^\mathrm{I-V}_i.
	\end{equation}
	These doppler and pseudocontinuum measurements are used in helioseismic investigations of the solar interior as well as in studies of photospheric convective flows and solar irradiance as a function of magnetic flux. We are interested in the magnetic field measurements, and will not consider them further here.

	Lastly, we note for future reference that the polarized spectra ($I_I(\lambda)$, $I_Q(\lambda)$, $I_U(\lambda)$, and $I_V(\lambda)$) seen by each of the pixels are integrated over the sky plane by the instrument PSF and by the pixel `box' function. Thus all of these measurements are made from area-averaged (or, equivalently, area-integrated) quantities, and should be treated as such. Therefore the magnetograms are more akin to fluxes than to point-wise magnetic field measurements.


\section{End-to-end measurement simulation}\label{sec:endtoend}


	The calibration is based on an end-to-end simulation of the measurement process. By this we mean that the simulation begins with the conditions where the light is emitted on the sun, continues to its collection by the telescope and detection by the sensors, and ends with its processing to form a magnetogram. The two `ends' of the end-to-end simulation are therefore the photosphere (as it is represented by the 3D MHD simulation datacube) and the magnetogram (as would be produced by the standard GONG processing). We now describe each step of this simulation process. The calibration curves will be based on the two endpoints of this simulation process, and their creation is described in the subsequent two papers \citep{PlowmanEtal_2019II, PlowmanEtal_2019III}. 
	It is not part of the end-to-end simulation, however, because many steps of the forward simulation lose information and in general the process is not reversible.

	\subsection{3D MHD Simulation \& Radiative Transfer} \label{sec:MHD_radtransfer}

\begin{figure}
		\includegraphics[width=0.5\textwidth]{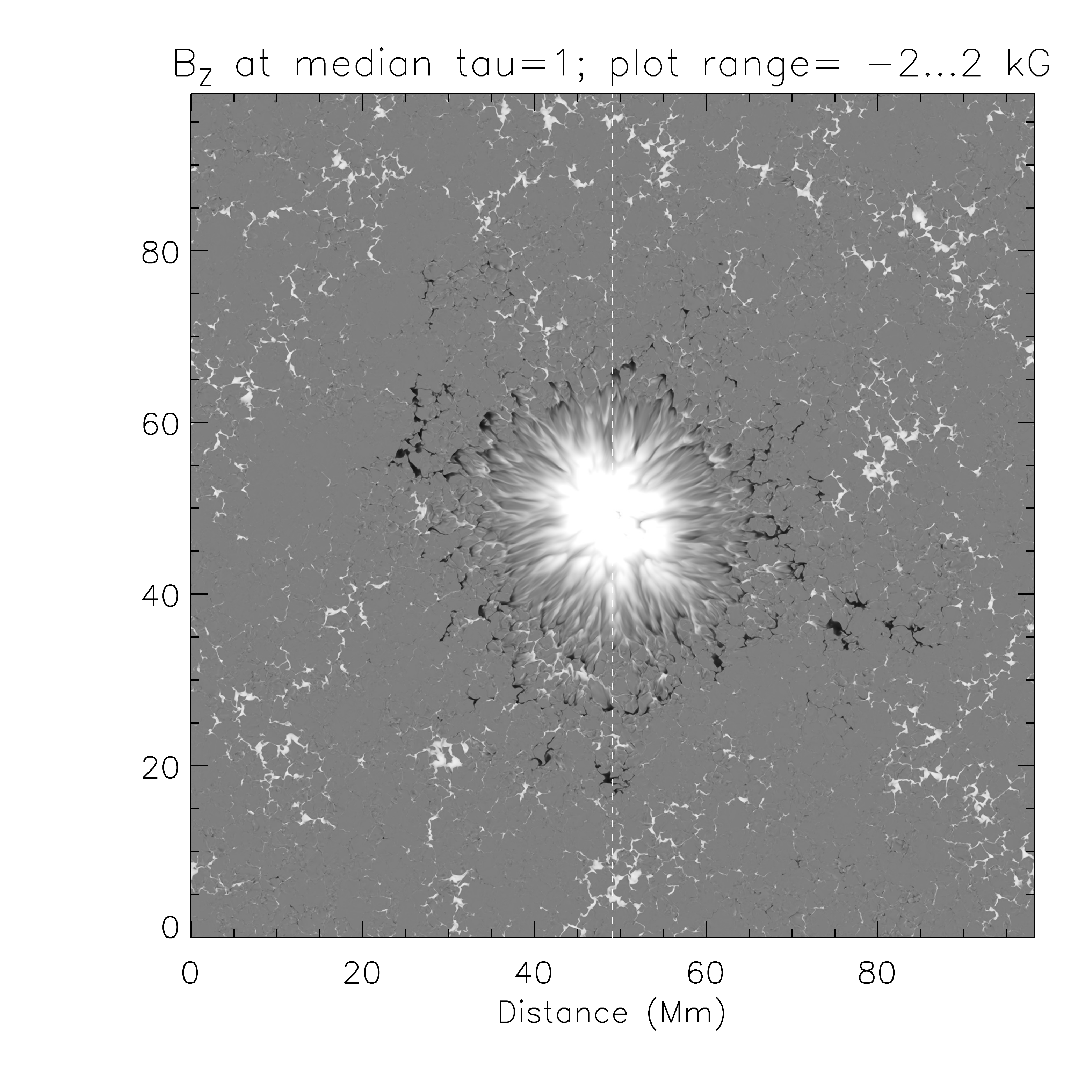}\includegraphics[width=0.5\textwidth]{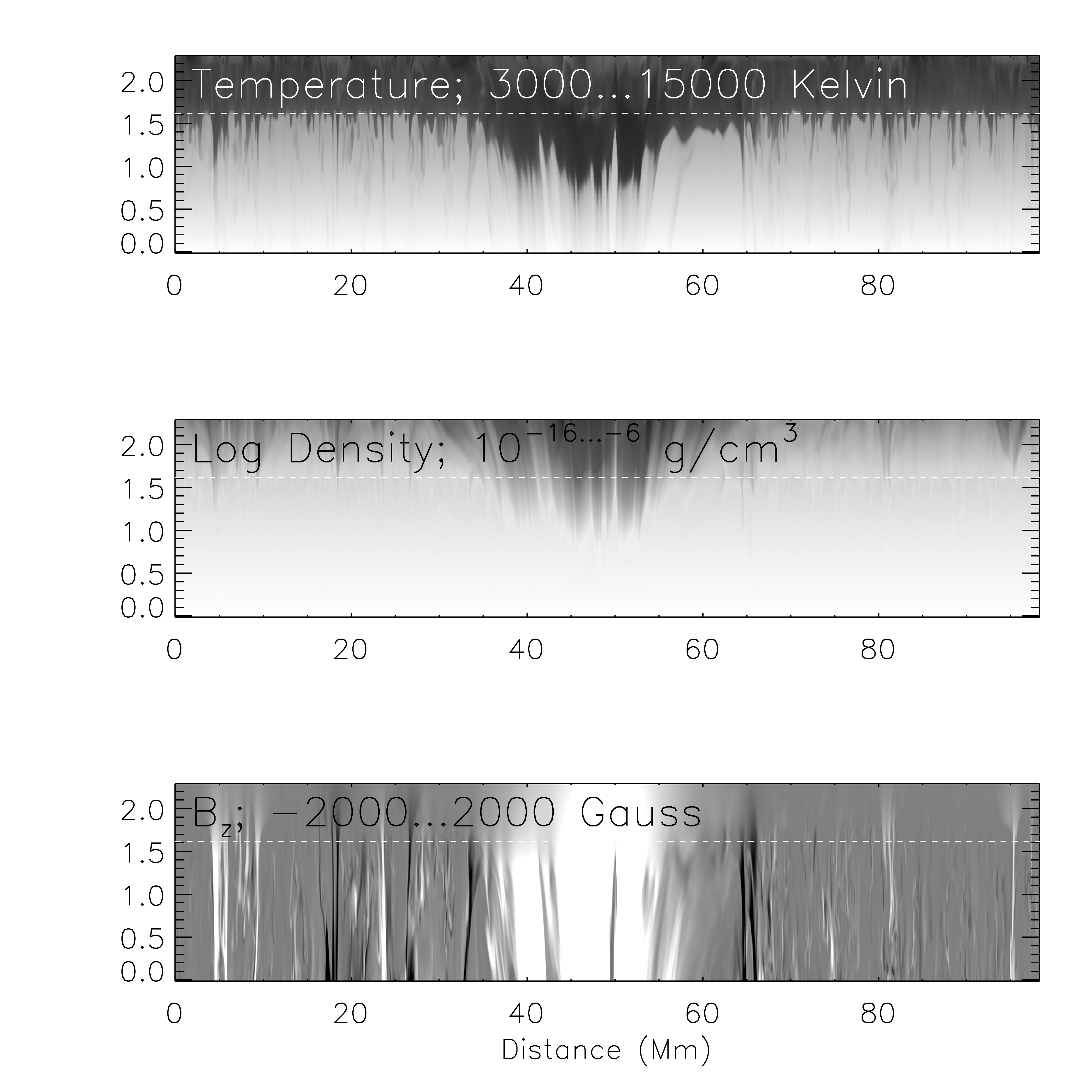}
		\caption{Horizontal (left) and vertical (right) slices of 3D MHD MURaM simulation from Mattias Rempel. Dashed line on left shows location of slices on right, and vice versa.}\label{fig:MHD_snapshot}
	\end{figure}

	For our 3D MHD `ground truth' simulation, we use a MURaM snapshot provided by Mattias Rempel. This snapshot is the simulation with penumbra discussed in \cite{Rempel2015}. It features a sunspot and its environs (the horizontal extent of the cube being roughly twice the diameter of the sunspot), providing both `Active Region' (sunspot) and `Quiet Sun' (non-sunspot) in the same simulation volume. The grid spacing of the snapshot is $48\times48$~km in the horizontal and 24~km in the vertical, while its dimensions are $2048\times2048$ volume elements horizontally and 96 in the vertical. The horizontal boundaries of the simulation wrap around, so that the top of the simulation box is identified with the bottom, and the left with the right. The simulation has a net positive magnetic polarity, and the horizontal flux at the top boundary has been enhanced (compared to a potential field) to produce a penumbra. Figure \ref{fig:MHD_snapshot} shows some representative slices.

	GONG measures the solar magnetic field via the intensity and polarization of solar light at specific wavelengths (as discussed in Section \ref{sec:GONGprinciple}). Producing these solar spectra from the 3D MHD snapshot requires simulating the detailed radiative transfer process by which the light is generated at and escapes from the solar atmosphere. For this, we use the Rybicki \& Hummer (RH) radiative transfer code developed by Han Uitenbroek \citep{Uitenbroek_ApJ2001}. The code is run in local thermodynamic equilibrium (LTE) mode at the full 48x48x24km resolution of the snapshot. We focus on the GONG Nickel line, sampling the spectra at 51 wavelengths ranging from 676.73 to 676.83 nm. This line can be blended with solar molecular lines in sunspots, and we do not include these lines; however, our focus in this calibration will be the environs of the sunspot more than the sunspot itself.
	
	To model the effects of inclination, we run the radiative transfer simulation at 6 inclined ray angles corresponding to 0, 25, 45, 60, 70, and 75 degrees of latitude (all at zero degree longitude). Due to memory usage, the snapshot is split into 32 vertical strips of 64x2048x96 volume elements (by using the full strip vertically we preserve the boundary wrapping property of the MHD simulation in the inclined ray direction, which is necessary for RH). These are combined after their spectra are computed, and binned down so that the final output spectra are 1024x1024 pixels with a resolution of 96x96 km. This is also done to limit memory and storage requirements, and is still much higher than the plate scale of synoptic magnetographs (HMI at $\sim 360$ km, GONG at $\sim 1800$ km). 

\begin{figure}
		\begin{center}\includegraphics[width=\textwidth]{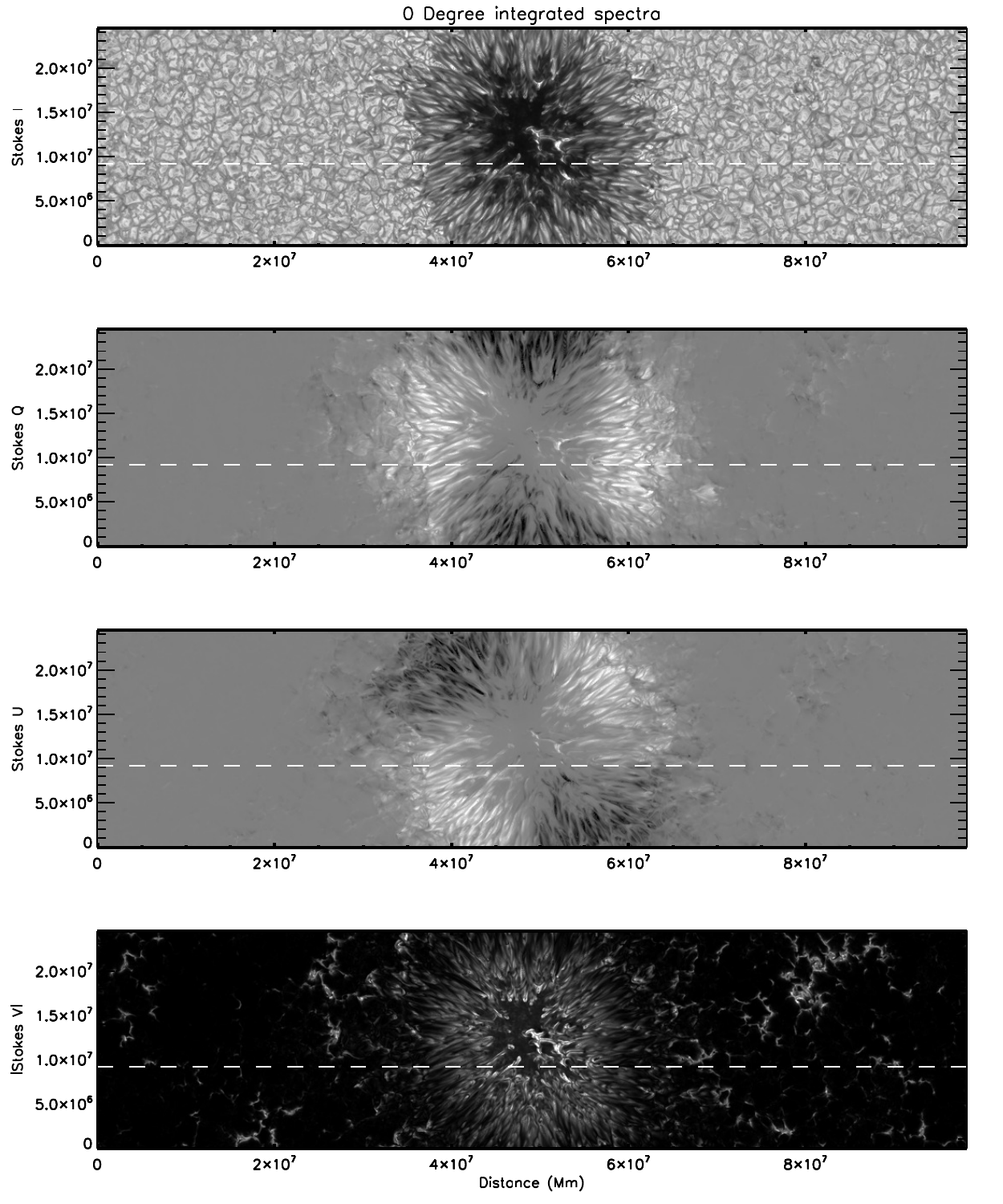}\end{center}
		\caption{GONG Ni I 676.8 nm intensities computed from 3D MHD simulation using RH. Dashed line shows location of spectral slices in Figure \ref{fig:rh_spectrum_vertical_spec_example}. This figure is for a vertical viewing angle (i.e., with the region at disk center).}\label{fig:rh_spectrum_vertical_inten_example}
	\end{figure}

\begin{figure}
		\begin{center}\includegraphics[width=\textwidth]{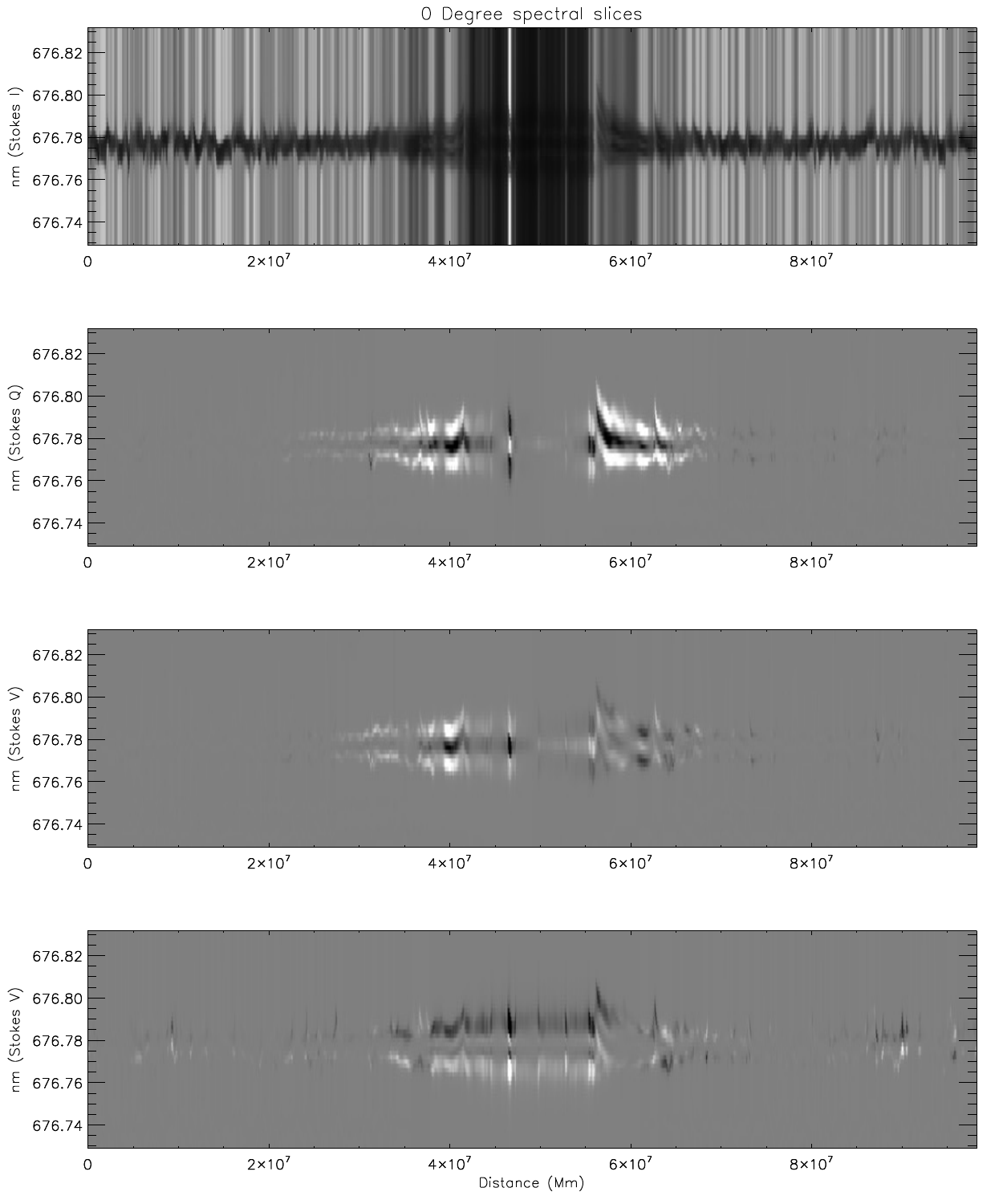}\end{center}
		\caption{GONG Ni I 676.8 nm spectral slice computed from 3D MHD simulation using RH. Dashed line in Figure \ref{fig:rh_spectrum_vertical_inten_example} shows the location of these spectral slices. This figure is for a vertical viewing angle (i.e., with the region at disk center).}\label{fig:rh_spectrum_vertical_spec_example}
	\end{figure}

\begin{figure}
		\begin{center}\includegraphics[width=\textwidth]{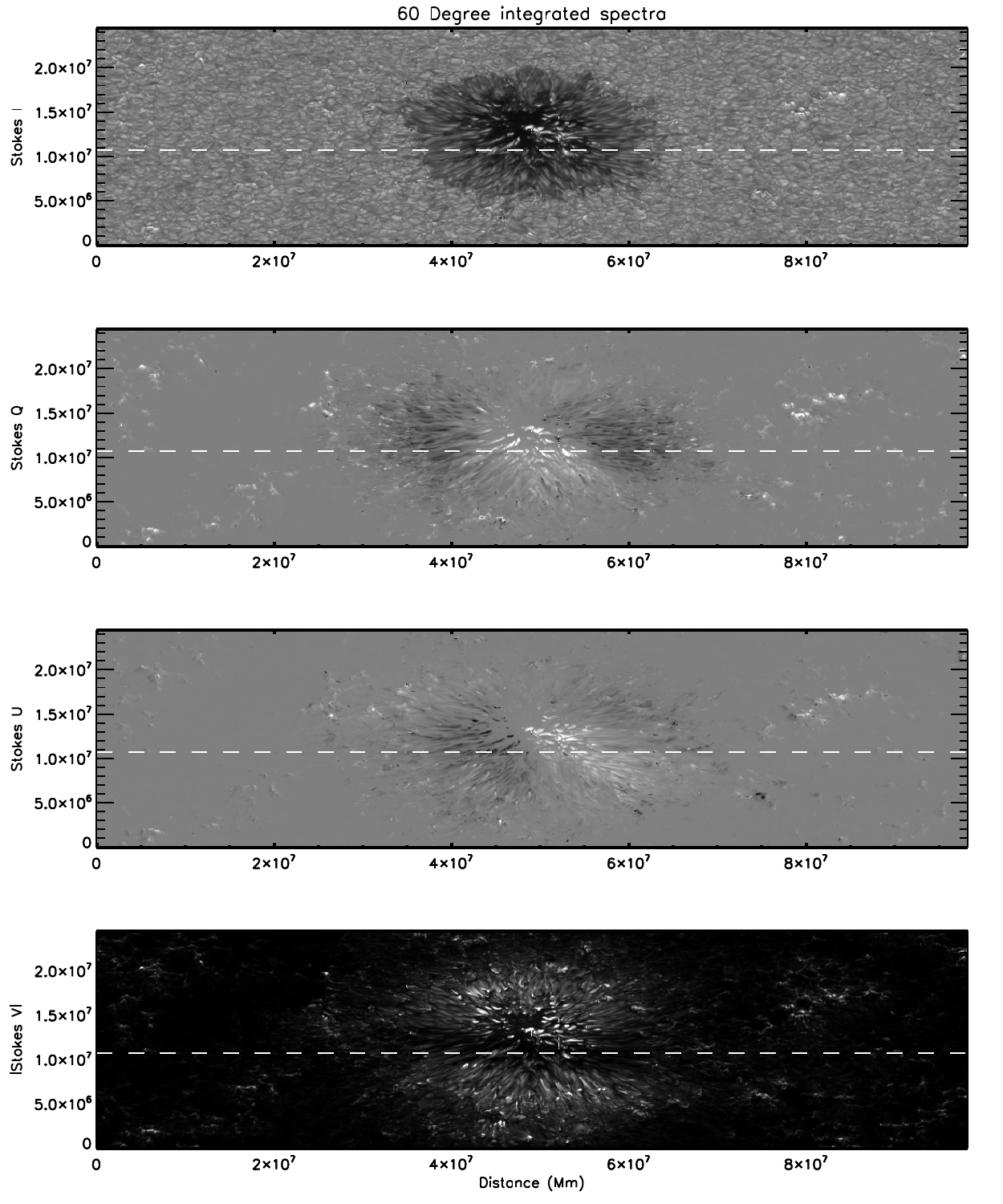}\end{center}
		\caption{GONG Ni I 676.8 nm intensities computed from 3D MHD simulation using RH. Dashed line shows location of spectral slices in Figure \ref{fig:rh_spectrum_60degree_spec_example}. This figure is for a 60 degree viewing angle (i.e., with the region at 60 degrees below disk center).}\label{fig:rh_spectrum_60degree_inten_example}
	\end{figure}

\begin{figure}
		\begin{center}\includegraphics[width=\textwidth]{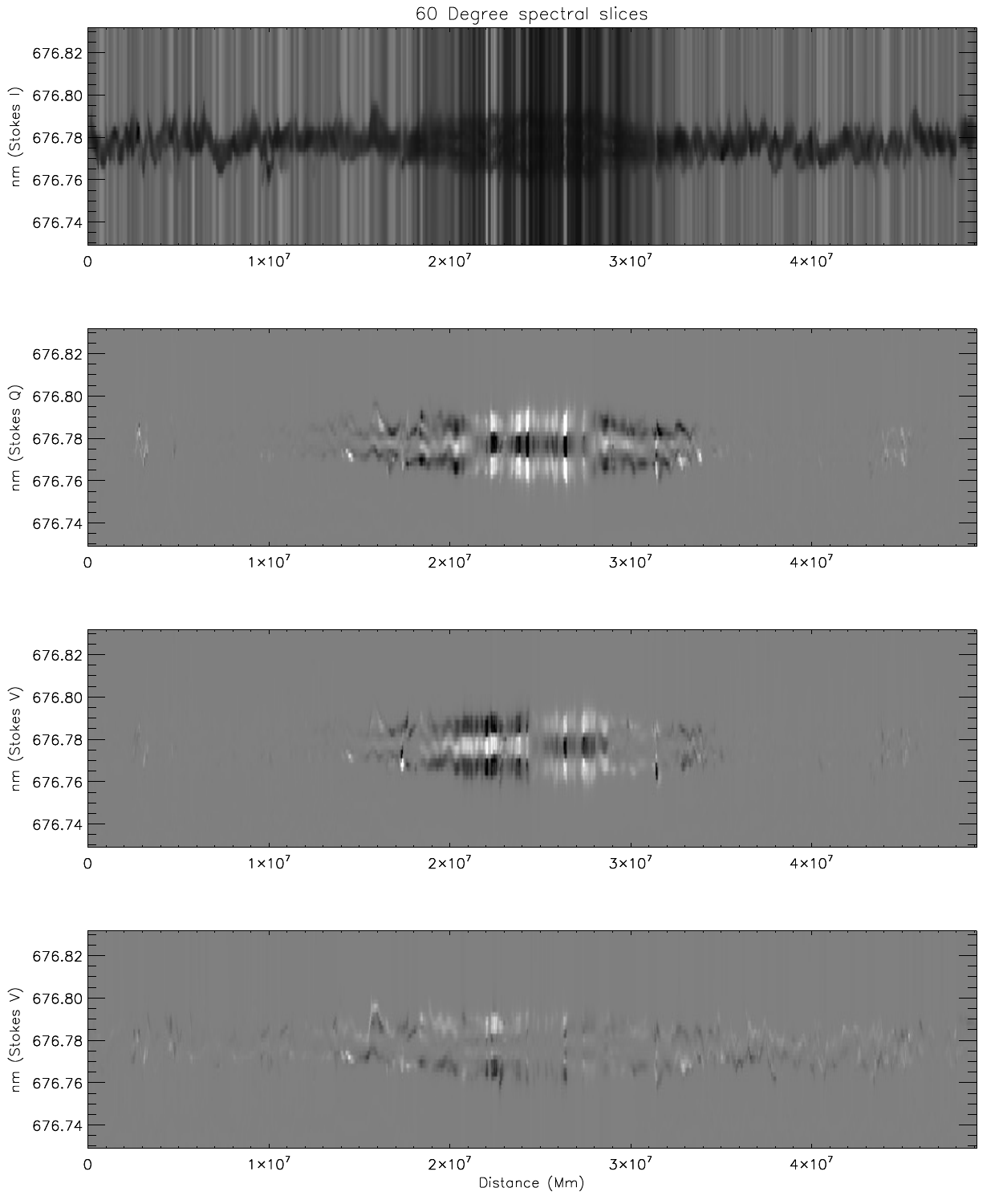}\end{center}
		\caption{GONG Ni I 676.8 nm spectral slice computed from 3D MHD simulation using RH. Dashed line in Figure \ref{fig:rh_spectrum_60degree_inten_example} shows the location of these spectral slices. This figure is for a 60 degree viewing angle (i.e., with the region at 60 degrees below disk center).}\label{fig:rh_spectrum_60degree_spec_example}
	\end{figure}

	Figures \ref{fig:rh_spectrum_vertical_inten_example} and \ref{fig:rh_spectrum_vertical_spec_example} illustrates the spectra viewed from the vertical, while Figures \ref{fig:rh_spectrum_60degree_inten_example} and \ref{fig:rh_spectrum_60degree_spec_example} shows the spectra at 60 degrees.

	\subsection{The GONG Simulator} \label{sec:GONGsimulator}

	This section describes the simulation of the GONG instrument, based on the understanding laid out in Section \ref{sec:GONGprinciple}. Our approach is to simulate the major observational effects, from solar emission to the detector plane to the final processed result, in the order they occur. Producing a precise optical model (describing every interface in the instrument) would be far too involved to be practical for this project, so we have settled for an intermediate approach: reproducing the most important effects in a way that's straightforward to program while leaving it to the processing to remove those effects which are easy to remove in that way. Thus, our simulation represents the nominal GONG, rather than an exact facsimile of a specific GONG instrument. To ensure clarity and reproducibility, the instrument simulation is now described step-by-step.

	The inputs to the instrument simulator are spatially resolved spectra for each stokes component (dimensions $n_x\times n_y\times n_\lambda$ for each of Stokes I, Q, U, and V). The linear polarization is not currently used, however: we restrict attention to the circular polarization states and we assume that there is no crosstalk, so the polarization states being measured are just I+V and I-V.

\begin{figure}
		\begin{center}\includegraphics[width=0.49\textwidth]{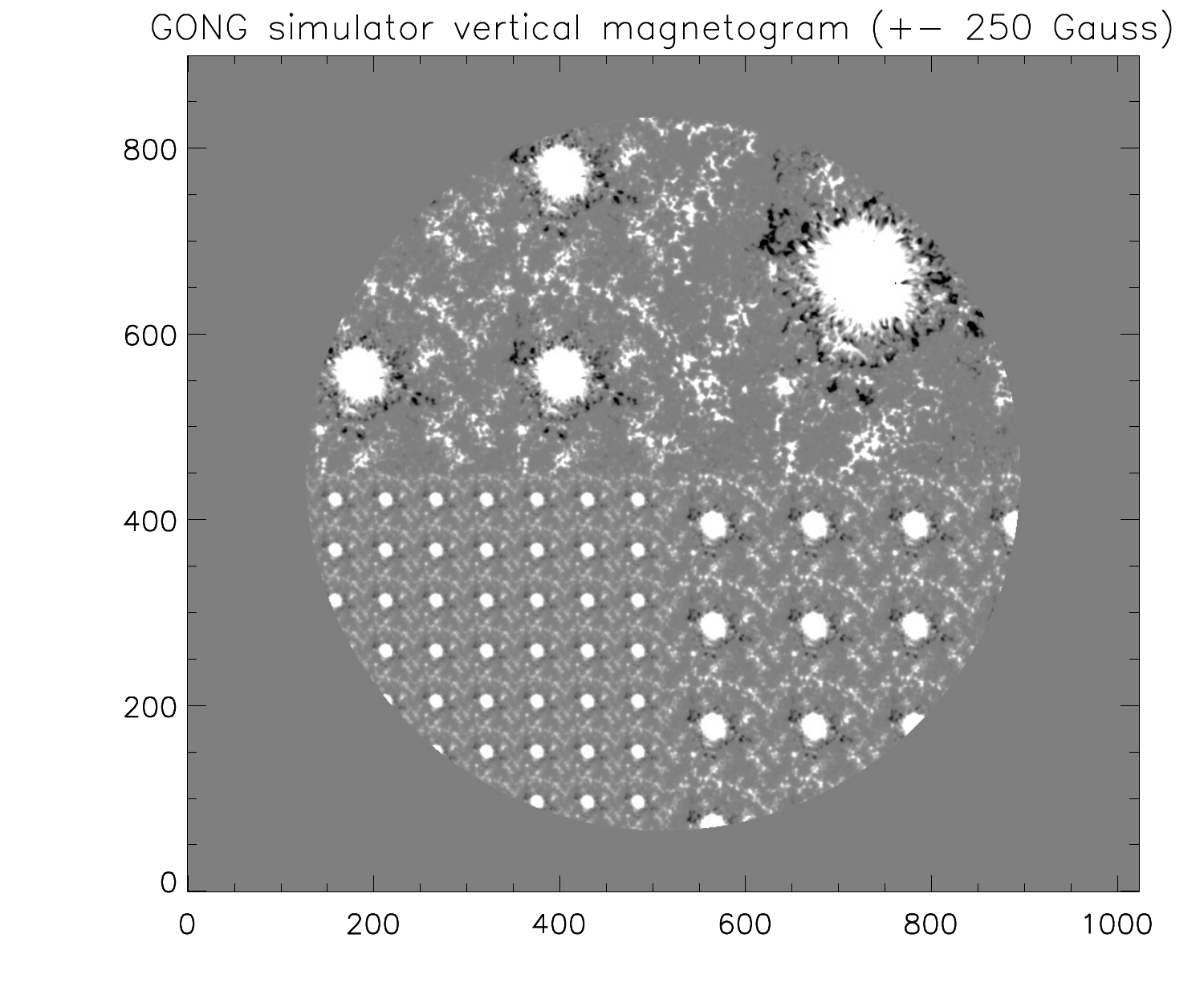}\includegraphics[width=0.49\textwidth]{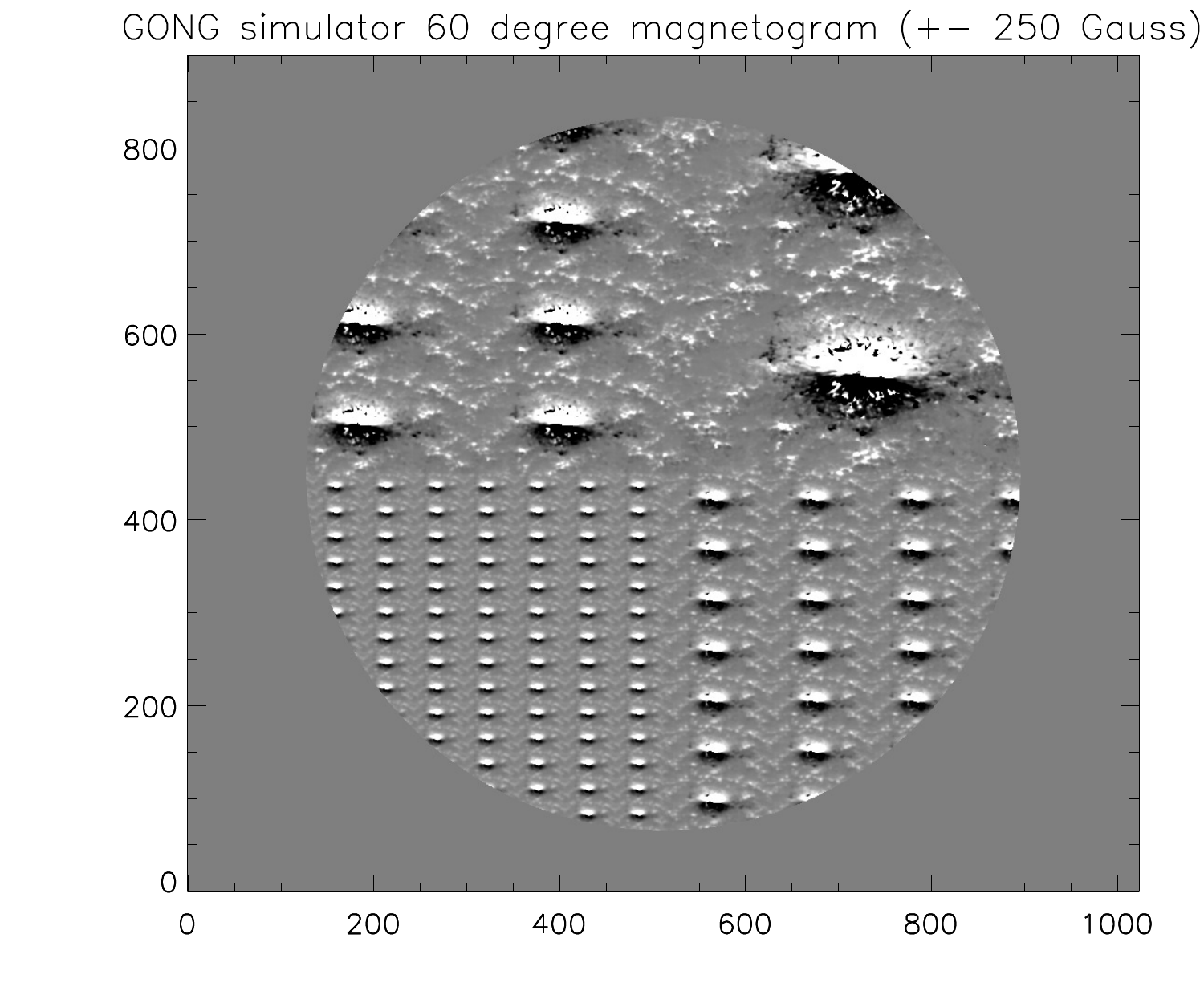}\end{center}
		\caption{Illustration of MURaM simulation tiling, and how multiple resolutions and different viewing angles are treated in the instrument simulation. Viewing angles are treated as separate images (left: 0 degrees, right: 60 degrees), different resolutions are used in each quadrant of the image (bottom left: GONG resolution, bottom right: twice GONG resolution, etc); Compare with Figure \ref{fig:MHD_snapshot}, which shows a single instance of the snapshot with no tiling or instrumental effects.}\label{fig:gongsimbz_tileref}
	\end{figure}

	These spectral cubes are produced from MURaM simulations by the RH radiative transfer code, as described in Section \ref{sec:MHD_radtransfer}. The MURaM simulations are much smaller than the whole sun due to computational limitations, so we tile them at several scales to cover the GONG detector plane. The intention is to increase the number of points in the calibration (via the tiling), and have some ability to investigate the effect of instrument resolution built into the simulator results. The effect of different viewing angles are investigated with an independent run of the simulator (resulting in a separate image), rather than tackling the thorny problem of stitching the simulation over a sphere. The tiling, and multiple resolutions, and separate images for each viewing angle are illustrated in Figure \ref{fig:gongsimbz_tileref}
	
	The result is a set of datacubes containing spectral radiance at the sun (in units of, e.g., $J s^{-1} m^{-2} Hz^{-1} Sr^{-1}$) in each of Stokes I, Q, U, and V.  The whole-disk tiling is is rebinned to $4096 \times 4096$ pixels, resulting in $\sim 5$ times GONG's plate scale; this reduces memory usage while still being high enough to avoid sampling artifacts.

	From this point, there are two phases to the simulation. The first is a setup phase:

	\begin{itemize}
		\item Convert the intensities produced by the radiative transfer simulation to spectral radiance received on Earth (i.e., erg $s^{-1} \mathrm{cm}^{-3}\mathrm{arcsecond}^{-2}$). That is, power received per unit aperture and wavelength (cm) per square arcsecond patch on the sky.
		\item Apply Mueller matrices to convert from individual Stokes components to Stokes I+V and I-V.
		\item Apply the instrument Point-Spread function (PSF). This is in two parts: an Airy function with a size of 6.1 arcseconds (See Section \ref{sec:GONGprinciple}), representing diffraction, and a 2-dimensional Gaussian function representing atmospheric seeing. GONG seeing is monitored as part of the data taking, and varies from site to site -- at some, the width of the atmospheric blurring is close to 2 arcseconds on average, while for others it is around 4 arcseconds. We use a width of 3.6 arcseconds for our Gaussian seeing kernel. In the real world, the seeing can also vary over the image plane and with time. This is not currently incorporated in our simulations.
		\item Pixelize the image, converting the spectral radiances to spectral fluxes per pixel (i.e., erg $s^{-1} \mathrm{cm}^{-3}$). This is done by first resampling the tiled whole-disk cube to 5 times the GONG resolution using bilinear interpolation and binning the result down to the GONG pixel size. As a result, artifacts from the interpolation will tend to be much smaller than the GONG pixel scale.
		\item Precompute the spectral fluxes modulated by each of the GONG filter response functions. These are split into two sets of 12 such functions (one for each polarization state) over 360 degrees, corresponding to each 30 degree rotation of the wave plate (since the pattern repeats every 90 degrees, 9 of these 12 functions are redundant, as described in Section \ref{sec:GONGprinciple}). To allow for coarser-grained wavelength effects, the integration is split into several wavelength subintervals, although currently no such effects are implemented. The result is the average spectral flux incident on the detector for each of the 12 waveplate angles and each wavelength subinterval. If the wavelength range of the spectral fluxes is less than the range of the prefilter, the wavelengths are padded with the continuum (edge).
		\item Multiply by GONG's effective area, assumed to be 0.6 $\mathrm{cm}^2$. This value is chosen so that simulated photon counts match those of the real GONG observations.
		\item Set the initial ADU offset to dithering (32 ``data number" units). The dithering is a feature of the real GONG data acquisition; it adds a repeating triangle wave to the analog signal read from the CCDs at each pixel and is intended to address some numerical artifacting in the discretization.
	\end{itemize}

	The second phase is time integration, accumulating frames as the waveplate rotates and the LCVR switches modes.
	Frames are accumulated onto 6 image planes (output images), one for each unique wavelength tuning and polarization state. These planes are meant to be equivalent to the 6 planes of a GONG raw file, although for simplicity we accumulate 10 minutes rather than averaging 10 separate one-minute magnetograms (as is the case for the GONG 10-minute magnetograms). This phase proceeds as follows:

	\begin{itemize}
		\item Each frame corresponds to a 1/60 second interval with a given LCVR state, nominal waveplate orientation, and one of the precomputed spectral fluxes described above.
		\item Dithering triangle wave: If the frame number is a multiple of 12 and we are in the first half of the current 1 minute interval, increment the dithering offset by two.
		\item Dithering triangle wave: If the frame number is a multiple of 12 and we are in the second half of the current 1 minute interval, decrement the dithering offset by two.
		\item Determine which of the 24 precomputed spectral fluxes applies to the current frame number.
		\item Compute, for the current spectral flux, the average detector count at each pixel in each of the wavelength subintervals for the given exposure time ($1/60$ second). The quantum efficiency of the GONG CCDs is 16\% .
		\item Multiply the mean detector counts by the flat field (taken from a real GONG flat field).
		\item Generate random detector counts from the mean detector counts according to the Poisson distribution.
		\item Add Gaussian read noise (assumed to be 4.8 ADU units) and Dark current (Poisson distributed based on a real GONG dark field).
		\item Divide by the ADU conversion factor (33) and round to the nearest integer.
		\item Add the current value of the dither offset.
		\item Add the resulting frame to the relevant image plane (i.e., in the output image array).
		\item Proceed to the next frame.	
	\end{itemize}

	Once all frames are finished, the three lowest bits are truncated, matching the treatment of the GONG data acquisition. GONG does this to ensure that the one-minute accumulated images fit into a 16-bit storage format without overflowing. Finally, the average dithering offset (approximately 182) is subtracted from the output image array.

	\subsection{Simulated Data Processing}

	Once the simulated raw data are produced they must be processed to obtain a magnetogram. We do not attempt to use the full GONG processing pipeline for this, because it is built to work on the full GONG operational data stream rather than the data set produced by the simulator. Instead, we use a simplified processing algorithm based on the same principles as the full GONG pipeline. It operates as follows:

	\begin{itemize}
		\item Subtract the dark frame (proportional to the number of frames).
		\item Divide by the flat (flats are normalized by their median, both at this processing stage and in the instrument simulator).
		\item Determine the Doppler shifts for I+V and I-V according to Equation \ref{eq:phase_dopplershift}.
		\item Determine each pixel's magnetic field measurement (average flux density) from the Doppler shifts (using Equation \ref{eq:BLOS}).
		\item Return the flux density and, optionally, the line average Doppler shift and pseudo-continuum intensity.
	\end{itemize}

	\noindent This is the final step of the GONG `end-to-end' instrument simulation process shown on the right side of Figure \ref{fig:calibration_boxdiagram}.

	\section{Results \& Summary}\label{sec:summary}

\begin{figure}
		\begin{center}\includegraphics[width=\textwidth]{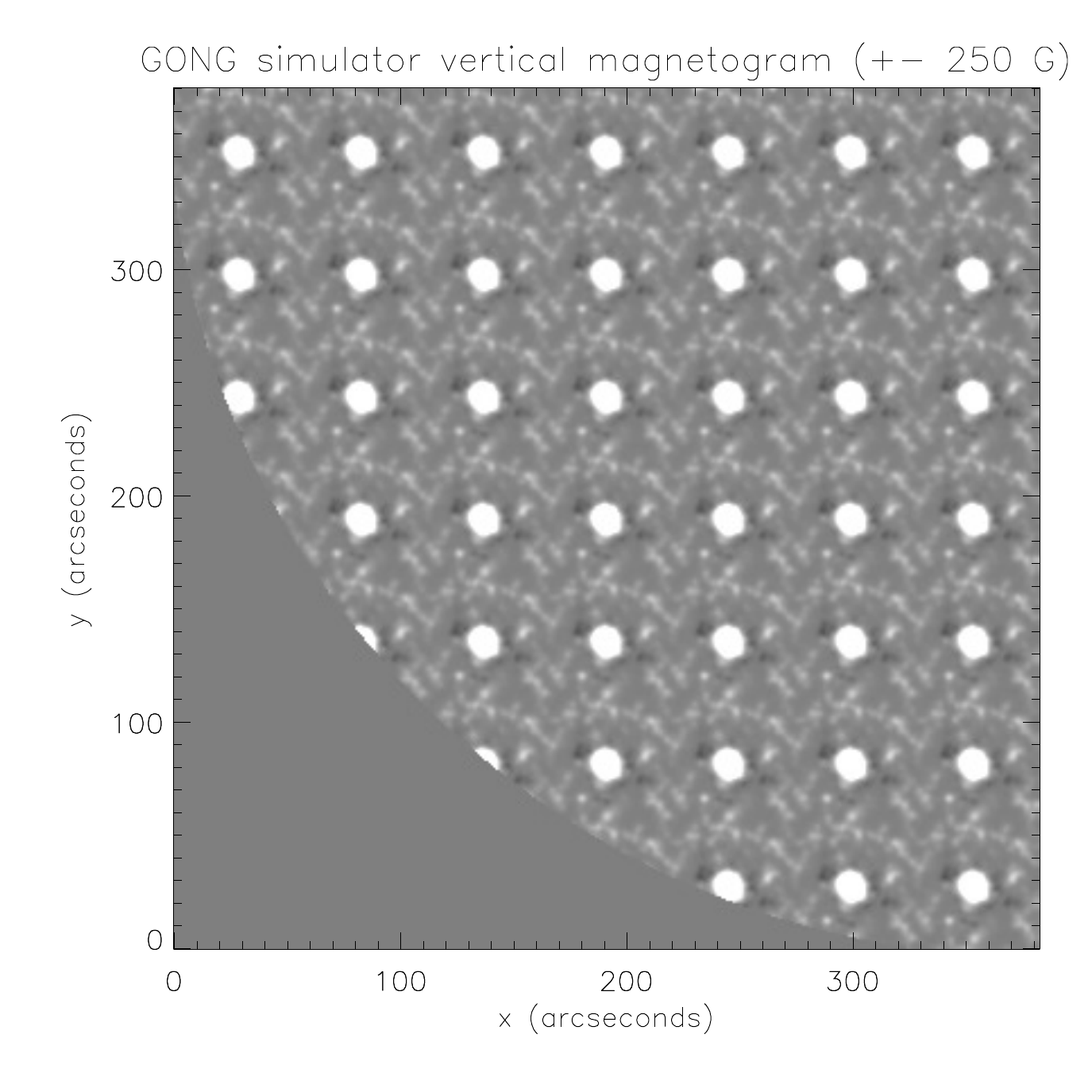}\end{center}
		\caption{Magnetograms resulting from GONG end-to-end simulation, for vertical viewing angle (i.e., with the region at disk center). Only the lower left of the simulated image is shown; compare with Figure \ref{fig:gongsimbz_tileref}.}\label{fig:gongsimbz00}
	\end{figure}

\begin{figure}
		\begin{center}\includegraphics[width=\textwidth]{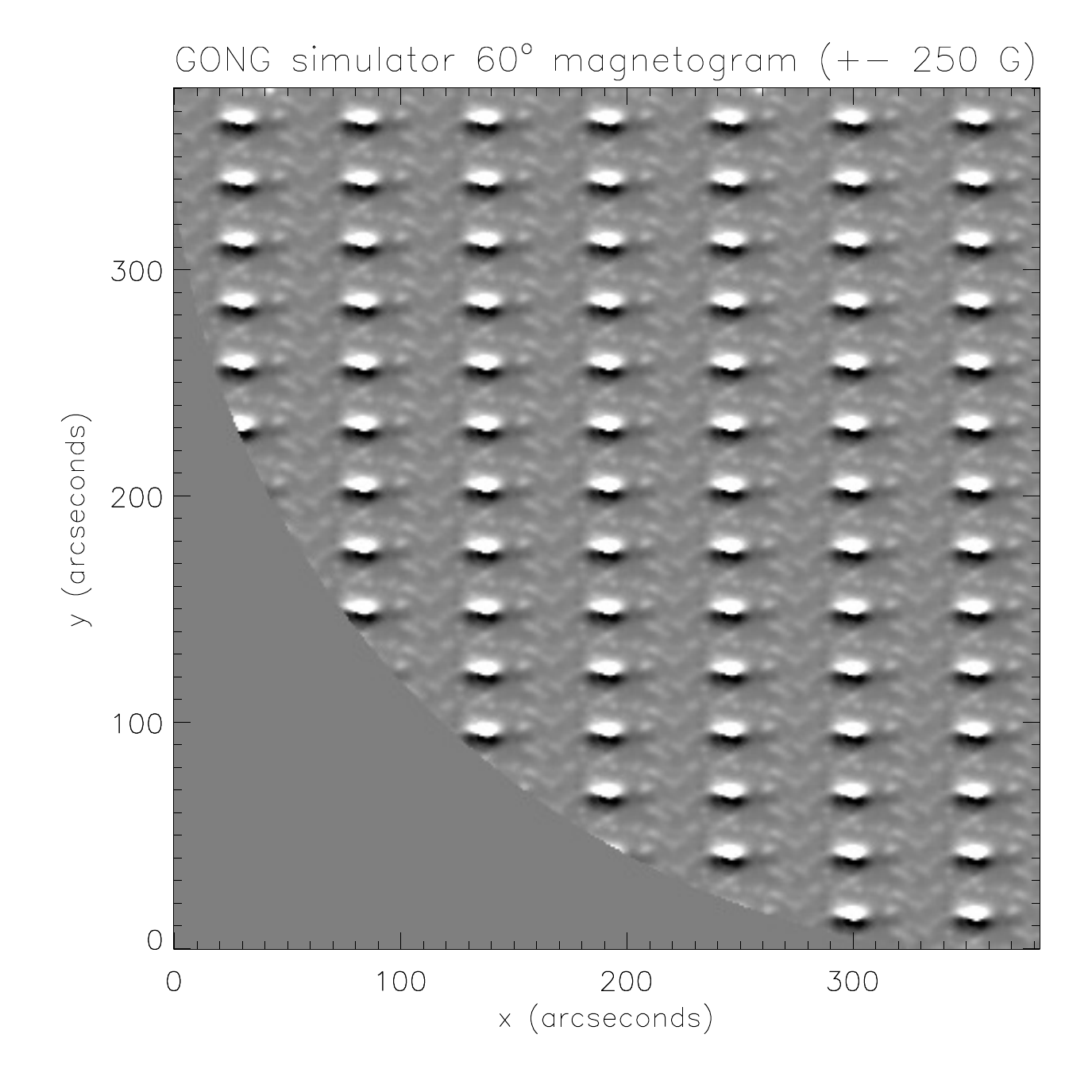}\end{center}
		\caption{Magnetograms resulting from GONG end-to-end simulation, for 60 degree viewing angle (i.e., at 60 degrees below disk center). Only the lower left of the simulated image is shown; compare with Figure \ref{fig:gongsimbz_tileref}.}\label{fig:gongsimbz60}
	\end{figure}

	Example magnetograms resulting from this end-to-end simulation were already shown, to illustrate our tiling of the MHD simulation, in Figure \ref{fig:gongsimbz_tileref}. However, we will largely be concentrating on the GONG-resolution portion of these simulation resolution (i.e., the lower left quadrant). These are shown in close-up in Figure \ref{fig:gongsimbz00}, for 0 degrees, and \ref{fig:gongsimbz60} for 60 degrees. We will defer close comparison of these with the MuRAM ground truth until we have discussed the ground truth `reduction', in the next paper \citep{PlowmanEtal_2019II}. 
	
	We also produce corresponding pseudo-continuum intensities and Doppler shifts; these are not shown for brevity. As previously mentioned, we also simulate 25, 45, 70, and 75 degrees; those likewise are not shown for brevity. Note that in this work we do not simulate the effects of longitude away from disk center -- only those of latitude. This is the most important longitude where synoptic maps are concerned, since the magnetic field values used to construct them are drawn from as close as possible to the central (i.e., Earth-directed) meridian. The results should be similar for all line-of-sight field measurements at a given inclination, however, and we treat our calibration curves as such.

	This concludes the description of our GONG instrument simulation and creation of synthetic magnetograms. We have also given an updated description of the GONG instrument, and we find (in Section \ref{sec:GONGprinciple}) that GONG is {\em not} subject to magnetograph saturation in the classical sense, at least up to strenths of $\sim 4$ kG. 
	Thus expressions such as that in \cite{WangSheeley95} are not directly relevant to GONG.

	The next paper in this series \citep{PlowmanEtal_2019II}, will consider the theory of magnetograph calibration and comparison in general, identifying common issues and developing methods to use the results of this simulation to create an `absolute' calibration for GONG. The resulting calibrations, and their implications, will be shown in a third, final, paper \citep{PlowmanEtal_2019III}.

	\subsection*{Acknowledgements}
	This work was funded in part by the NASA Heliophysics Space Weather Operations-to-Research program, grant number 80NSSC19K0005, and by a University of Colorado at Boulder Chancellor’s Office Grand Challenge grant for the Space Weather Technology, Research, and Education Center (SWx TREC).

	We acknowledge contributions, discussion, information, and insight from a variety of sources: Gordon Petrie, Jack Harvey, Valentin Martinez Pillet, Sanjay Gosain, and Frank Hill, among others.

	\section*{Disclosure of Potential Conflicts of Interest}
	The authors declare that they have no conflicts of interest.

	\bibliographystyle{apj}
	\bibliography{apj-jour,GONG_simulatorI}
\end{document}